\documentclass[runningheads]{svmult}

\usepackage{makeidx}   % allows index generation
\usepackage{graphicx}  % standard LaTeX graphics tool
\usepackage{subeqnar}  % subnumbers individual equations
\usepackage{multicol}  % used for the two-column index
\usepackage{physprbb}  % modified textarea for proceedings,
\makeindex             % used for the subject index
\begin{document}
\title*{Chaos or Order in Double Barred Galaxies?}
\toctitle{Chaos or Order in Double Barred Galaxies?}
% allows explicit linebreak for the table of content
%
%
\titlerunning{Chaos or Order in Double Barred Galaxies?}
% allows abbreviation of title, if the full title is too long
% to fit in the running head
%
\author{Witold Maciejewski}
\authorrunning{Witold Maciejewski}
% if there are more than two authors,
% please abbreviate author list for running head
%
%
\institute{INAF -- Osservatorio Astrofisico di Arcetri,
Largo E. Fermi 5, 50125 Firenze, Italy, \\ 
and Obserwatorium Astronomiczne Uniwersytetu Jagiello{\'n}skiego,
Krak{\'o}w, Poland}

\maketitle              % typesets the title of the contribution

%\today

%\baselineskip1cm

\begin{abstract}
Bars in galaxies are mainly supported by particles trapped around closed 
periodic orbits. These orbits respond to the bar's forcing frequency only and 
lack free oscillations. We show that a similar situation takes place in double 
bars: particles get trapped around orbits which only respond to the forcing 
from the two bars and lack free oscillations. We find that writing the 
successive
positions of a particle on such an orbit every time the bars align generates 
a closed curve, which we call a loop. Loops allow us to verify consistency
of the potential. As maps of doubly periodic orbits, loops can be used to
search the phase-space in double bars in order to determine the fraction 
occupied by ordered motions.
\end{abstract}

\section{Introduction}
Bars within bars appear to be a common phenomenon in galaxies. Recent
surveys show that up to 30\% of early-type barred galaxies contain nested bars
\cite{erw+s02}. The relative orientation of the two
bars is random, therefore it is likely that the bars rotate with different
pattern speeds. Inner bars, like large bars, are made of relatively old
stellar populations, since they remain distinct in near infrared 
\cite{wozn96}. Galaxies with two independently rotating bars do not conserve
the Jacobi integral, and it is a complex dynamical task to explain how such 
systems are sustained. To account for their longevity, one has to find sets 
of particles that support the shape of the potential in which they move.
Particle motion in a potential of double bars belongs to the general
problem of motion in a pulsating potential \cite{l+g89} \cite{sridh88},
of which the restricted elliptical 3-body problem is the best known
example. Families of closed periodic orbits have been found 
in this last problem, where the test particle moves 
in the potential of a binary star with components on elliptical orbits 
\cite{bro69}. However, such families are parameterized by values that also 
characterize the potential (i.e. ellipticity of the stellar orbit and the 
mass ratio of the stars), and their orbital periods are commensurate with
the pulsation period of the potential. For a given potential, 
these families are reduced to single orbits separated in phase-space. 
The solution for double bars is formally identical, and there an orbit can 
close only when the orbital period is commensurate with the relative 
period of the bars. Such orbits are separated in phase-space, and therefore
families of closed periodic orbits are unlikely to provide orbital support 
for nested bars. Another difficulty in supporting nested bars is caused by 
the piling up of resonances created by each bar, which leads to considerable 
chaotic zones. In order to minimize the 
number of chaotic zones, resonant coupling between the bars has been proposed
\cite{sygnet88}, so that the resonance generated by one bar overlaps with 
that caused by the other bar.

Finding support for nested bars has been hampered by the fact that closed
periodic orbits are scarce there. However, it is particles, not orbits, 
which create density distributions that support the potential. The concept 
of closed periodic orbit is too limiting in investigation of nested bars, and 
another description of particle motion, which does not have its limitations,
is needed. Naturally, in systems with two forcing frequencies, double-periodic 
orbits play a fundamental role. Thus in double bars a large fraction of 
particle trajectories gets trapped around a class of double-periodic orbits.
Although such orbits do not close in any reference frame, they can be 
conveniently mapped onto the loops \cite{m+s00}, which are an efficient 
descriptor of orbital structure in a pulsating potential. The loop is a
closed curve that is made of particles moving in the potential of a 
doubly barred galaxy, and which pulsates with the relative period of the bars.
Orbital support for nested bars can be provided by placing particles on
the loops. 

Here I give a systematic description of the loop approach, which recovers
families of stable double-periodic orbits, and which can be applied to any 
pulsating potential. In \S2 I use the epicyclic approximation to introduce
the basic concepts, and in \S3 I outline the general method.

\section{The epicyclic solution for any number of bars}
If a galaxy has a bar that rotates with a constant pattern speed,
it is convenient to study particle orbits in the reference frame rotating
with the bar. If two or more bars are present, and each rotates with 
its own pattern speed, there is no reference frame in which the potential
remains unchanged. In order to point out formal similarities in solutions 
for one and many bars, I solve the linearized equations in the inertial
frame. This is equivalent to the solution in any rotating frame, and the 
transformation is particularly simple: in the rotating frame the 
centrifugal and Coriolis terms are equivalent to the Doppler shift of the
angular velocity. It is convenient to show it in cylindrical coordinates 
$(R,\varphi,z)$: if ${\bf e}_z$ is the rotation axis, then the $R$ and 
$\varphi$ components of the equation of motion for the rotating frame,
$\ddot{\bf r} = -\nabla \Phi  - 2 ({\bf \Omega_B} \times \dot{\bf r})
- {\bf \Omega_B} \times ( {\bf \Omega_B} \times {\bf r})$, can be written as
\begin{eqnarray*}
\ddot{R} - R ( \dot{\varphi} + \Omega_B )^2
& = & -\frac{\partial \Phi}{\partial R}, \\
R \ddot{\varphi} + 2 \dot{R} ( \dot{\varphi} + \Omega_B )
& = & -\frac{1}{R} \frac{\partial \Phi}{\partial \varphi}.
\end{eqnarray*}
These equations are identical with the components of the equation
of motion in the inertial frame,
\begin{equation}
 \ddot{\bf r} = -\nabla \Phi ,
\label{eqmi}
\end{equation}
where clearly the angular velocity $\dot{\varphi}$ in the rotating 
frame corresponds to $\dot{\varphi}+\Omega_B$ in the inertial
frame. For the rest of this section I assume the inertial frame,
in which the equation of motion (\ref{eqmi}) has the following $R$ and 
$\varphi$ components in cylindrical coordinates
\begin{eqnarray}
\label{eqmri}
\ddot{R} - R \dot{\varphi}^2
& = & -\frac{\partial \Phi}{\partial R}, \\
\label{eqmfi}
R \ddot{\varphi} + 2 \dot{R} \dot{\varphi}
& = & -\frac{1}{R} \frac{\partial \Phi}{\partial \varphi} .
\end{eqnarray}
The $z$ component in any frame is $\ddot{z} =  - \partial \Phi / \partial z$,
but I consider here motions in the plane of the disc only, hence I neglect 
the dependence on $z$.

To linearize equations (\ref{eqmri}) and (\ref{eqmfi}), one needs
expansions of $R$, $\varphi$ and $\Phi$ to first order terms. The epicyclic
approximation is valid for particles whose trajectories oscillate
around circular orbits. For such particles one can write
\begin{eqnarray}
R(t)              & = & R_0 + R_I(t) , \\
\varphi(t)        & = & \varphi_{00} + \Omega_0 t + \varphi_I(t) , \\
\Phi(R,\varphi,t) & = & \Phi_0(R) + \Phi_I(R,\varphi,t) ,
\end{eqnarray}
where terms with index $I$ are small to the first order, and second- and 
higher-order terms were neglected. The parameter $\varphi_{00}$ allows the 
particle to start from any position angle at time $t=0$,
so that $\varphi_0 = \varphi_{00} + \Omega_0 t$. Asymmetry $\Phi_I$
in the potential is small and may be time-dependent. The angular velocity 
$\Omega_0$ on the circular orbit of radius $R_0$ relates to the 
potential $\Phi_0$ through the zeroth order of (\ref{eqmri}):
$\Omega_0^2 = (1/R_0) (\partial \Phi_0 / \partial R) \; |_{R_0}$.
The zeroth order of (\ref{eqmfi}) is identically equal to zero, and 
the first order corrections to (\ref{eqmri}) and (\ref{eqmfi}) take
respectively forms
\begin{eqnarray}
\label{rlin}
\ddot{R_I}  -  4 A \Omega_0 R_I  -  2 R_0 \Omega_0 \dot{\varphi_I}  & = &  
-\frac{\partial \Phi_I}{\partial R} \; |_{R_0,\varphi_0}, \\
\label{flin}
R_0 \ddot{\varphi_I}  + 2 \Omega_0 \dot{R_I}  & = &  
-\frac{1}{R_0} \frac{\partial \Phi_I}{\partial \varphi}  \; |_{R_0,\varphi_0},
\end{eqnarray}
where $A$ is the Oort constant defined by 
$ 4 A \Omega_0 = \Omega_0^2 - \frac{\partial^2 \Phi_0}{\partial R^2} |_{R_0}$.

We assume that the bars are point-symmetric
with respect to the galaxy centre. Thus to first order the departure
of the barred potential from axial symmetry can be described by a term
$\cos (2 \varphi)$. If multiple bars, indexed by $i$, rotate independently 
as solid bodies with angular velocities $\Omega_i$, the 
time-dependent first-order correction $\Phi_I$ to the potential can be
written as
\begin{equation}
\label{phi1}
\Phi_I(R,\varphi,t) = \sum_i \Psi_i(R) \cos[ 2 (\varphi - \Omega_i t) ] ,
\end{equation}
where the radial dependence $\Psi_i(R)$ has been separated from the angle 
dependence. No phase in the trigonometric functions above means that we 
define $t=0$ when all the bars are aligned. Derivatives of (\ref{phi1})
enter right-hand sides of (\ref{rlin}) and (\ref{flin}), which after
introducing $\omega_i = 2 (\Omega_0 - \Omega_i)$ take the form
\begin{eqnarray}
\label{rlin1}
\ddot{R_I}  -  4 A \Omega_0 R_I  -  2 R_0 \Omega_0 \dot{\varphi_I}  & = &  
- \sum_i \frac{\partial \Psi_i}{\partial R} \; |_{R_0} \cos(\omega_i t + 2 \varphi_{00}) , \\
\label{flin1}
R_0 \ddot{\varphi_I}  + 2 \Omega_0 \dot{R_I}  & = &  
\frac{2}{R_0} \sum_i \Psi_i(R_0) \sin (\omega_i t + 2 \varphi_{00}).
\end{eqnarray}

In order to solve the set of equations (\ref{rlin1},\ref{flin1}), one can 
integrate (\ref{flin1}) and get an expression for $R_0 \dot{\varphi_I}$, 
which furthermore can be substituted to (\ref{rlin1}). This substitution
eliminates $\varphi_I$, and one gets a single second order equation for $R_I$,
which can be written schematically as 
\begin{equation}
\label{ddotR}
\ddot{R_I} + \kappa_0^2 R_I = \sum_i A_i \cos (\omega_i t + 2 \varphi_{00})
                              + C_{\varphi},
\end{equation}
where 
$ A_i = - \frac{4 \Omega_0 \Psi_i}{\omega_i R_0}
        - \frac{\partial \Psi_i}{\partial R}_{| R_0} $,
$\kappa_0^2 = 4 \Omega_0 ( \Omega_0 - A )$, and $C_{\varphi}/2\Omega_0$ is
the integration constant that appears after integrating (\ref{flin1}). This 
is the equation of a harmonic oscillator with multiple forcing terms, whose
solution is well known. It can be written as
\begin{equation}
\label{r1}
R_I (t) = C_1 \cos (\kappa_0 t + \delta) 
   + \sum_i M_i \cos(\omega_i t + 2 \varphi_{00})
   + C_{\varphi}/\kappa_0^2 .
\end{equation}
The first term of this solution corresponds to a free oscillation
at the local epicyclic frequency $\kappa_0$, and $C_1$ is unconstrained.
The terms under the sum describe oscillations resulting from the forcing 
terms in (\ref{phi1}), and $M_i$ are functions of $A_i$. Hereafter I focus 
on solutions without free oscillations, thus I assume that $C_1=0$. These
solutions will lead to closed periodic orbits and to loops. The formula for 
$\varphi_I(t)$ can be obtained by substituting (\ref{r1}) into the 
time-integrated (\ref{flin1}). As a result, one gets
\begin{equation}
\label{phi1dot}
\dot{\varphi_I} =  \sum_i N_i \cos(\omega_i t + 2 \varphi_{00})
                 - \frac{2 A C_{\varphi}}{\kappa_0^2 R_0},
\end{equation}
where again $N_i$ are determined by the coefficients of the 
equations above. Note that to the first order
$\Omega_0 [R_0 + C_{\varphi}/\kappa_0^2] = 
\Omega_0 [R_0] - 2 A C_{\varphi}/{\kappa_0^2 R_0}$,
thus the integration constants entering (\ref{r1}) and
(\ref{phi1dot}) correspond to a change in the guiding radius $R_0$, and to
the appropriate change in the angular velocity $\Omega_0$. They all can
be incorporated into $R_0$, and in effect the unique solutions for $R_I$ and
$\varphi_I$ are
\begin{eqnarray}
\label{r1f}
R_I (t)         & = & \sum_i  M_i \cos(\omega_i t + 2 \varphi_{00}) , \\
\label{f1f}
\varphi_I(t)    & = & \sum_i N'_i \cos(\omega_i t + 2 \varphi_{00}) + const ,
\end{eqnarray}
where free oscillations have been neglected. The integration constant in
(\ref{f1f}) is an unconstrained parameter of the order of $\varphi_I$.

\subsection{Closed periodic orbits in a single bar}
In a potential with a single bar there is only one term in the sums 
(\ref{r1f}) and (\ref{f1f}), hereafter indexed with $B$. 
Consider the change in values of $R_I$ and 
$\varphi_I$ for a given particle after half of its period in the frame 
corotating with the bar. This interval is taken because the bar is 
bisymmetric, so its forcing is periodic with the period $\pi$ in angle. 
After replacing $t$ by $t + \pi / (\Omega_0 - \Omega_B)$ one gets
\begin{eqnarray*}
R_I & = &
M_B \cos[ \omega_B (t + \frac{\pi}{\Omega_0 - \Omega_B}) + 2 \varphi_{00}] \\
& = & 
M_B \cos( \omega_B t + 2 \pi + 2 \varphi_{00} ).
\end{eqnarray*}
Thus the solution for $R_I$ after time $\pi / (\Omega_0 - \Omega_B)$
returns its starting value, and the same holds true for $\varphi_I$.
After twice that time, i.e. in a full period of this particle in the 
bar frame, the epicycle centre returns to its starting point and the
orbit closes. Thus (\ref{r1f}) and (\ref{f1f}) describe closed periodic
orbits in the linearized problem of a particle motion in a single bar.

\subsection{Loops in double bars}
When two independently rotating bars coexist in a galaxy (hereafter indexed
by $B$ and $S$), there is no reference frame in which the potential is 
constant. Thus when a term
from one bar in (\ref{r1f}) and (\ref{f1f}) returns to its starting value,
the term from the other bar does not (unless the frequencies of the bars
are commensurate). Therefore the particle's trajectory does not close in 
any reference frame. However, consider the change in value of $R_I$ and 
$\varphi_I$ after time $\pi / (\Omega_S - \Omega_B)$, which is the relative
period of the bars. One gets
\begin{eqnarray*}
R_I & = &
M_B \cos[ \omega_B (t + \frac{\pi}{\Omega_S - \Omega_B}) + 2 \varphi_{00}] +
M_S \cos[ \omega_S (t + \frac{\pi}{\Omega_S - \Omega_B}) + 2 \varphi_{00}] \\
& = & 
M_B \cos( \omega_B t + 2 \pi \frac{\Omega_0 - \Omega_B}{\Omega_S - \Omega_B} + 2 \varphi_{00} ) +
M_S \cos( \omega_S t + 2 \pi \frac{\Omega_0 - \Omega_S}{\Omega_S - \Omega_B} + 2 \varphi_{00} ) \\
& = & 
M_B \cos( \omega_B t + 2 \pi + 2 \varphi_{01} ) +
M_S \cos( \omega_S t + 2 \varphi_{01}),
\end{eqnarray*}
where $\varphi_{01} = \varphi_{00} + \pi \frac{\Omega_0 - \Omega_S}{\Omega_S - \Omega_B}$. The same result can be obtained for $\varphi_I$. This means that
the time transformation $t \rightarrow t + \pi / (\Omega_S - \Omega_B)$ is 
equivalent to the change in the starting position angle of a particle from
$\varphi_{00}$ to $\varphi_{01}$. Consider motion of a set of particles 
that have the same guiding radius $R_0$, but start at various position 
angles $\varphi_{00}$. This is a one-parameter set, therefore in the disc 
plane it is represented by a curve, and because of continuity of 
(\ref{r1f}) and (\ref{f1f}) this curve is closed. After time 
$\pi / (\Omega_S - \Omega_B)$, a particle starting at angle $\varphi_{00}$ will
take the place of the particle which started at $\varphi_{01}$, a particle 
starting at $\varphi_{01}$ will take the place of another particle from this
curve and so on. The whole curve will regain its shape and position every
$\pi / (\Omega_S - \Omega_B)$ time interval, although positions of particles
on the curve will shift. This curve is the epicyclic approximation to the 
{\it loop}: a curve made of particles moving in a given potential, such that 
the curve returns to its original shape and position periodically. In the case
of two bars, the period is the relative period of the bars, and the loop
is made out of particles having the same guiding radius $R_0$. Particles
on the loop respond to the forcing from the two bars, but they lack any 
free oscillation. An example of a set of loops in a doubly barred galaxy 
in the epicyclic approximation can be seen in \cite{m+s97}. Since they 
occupy a significant part of the disc, one should anticipate large
zones of ordered motions also in the general, non-linear solution for
double bars.

\section{Full nonlinear solution for loops in nested bars}
Tools and concepts useful in the search for ordered motions in
double bars are best introduced through the
inspection of particle trajectories in such systems. For this inspection
I chose the potential of Model 1 defined in \cite{m+s00}, where the small
bar is 60\% in size of the big bar, and pattern speeds of the bars
are not commensurate. Consider a particle moving in this potential
inside the corotation of the small bar. Simple experiments with
various initial velocities show that if the initial velocity is 
small enough, the particle usually remains bound. A typical trajectory 
is shown in the left panels of Fig.1 -- since it depends on the reference 
frame, it is written twice, for reference frame of each bar. Further 
experimenting with initial velocities shows that particle trajectories are 
often even tidier: they look like those in the right
panels of Fig.1, as if the trajectories were trapped around some regular
orbit. 

\begin{figure}[b]
\vspace{-5mm}
\begin{center}
\includegraphics[width=125mm]{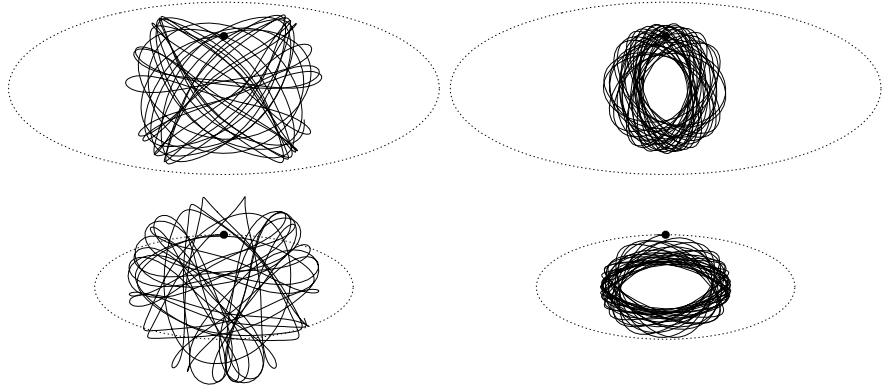}
\end{center}
\vspace{-12cm}
\caption[]{Two example trajectories (one in the two left panels, one in 
the two right ones) of a particle that moves in the
potential of two independently rotating bars. The particle is followed 
for 10 relative periods of the bars, and its trajectory is displayed in 
the frame corotating with the big bar (top panels), and the small bar 
(bottom panels). 
Each bar is outlined in its own reference frame by the dotted line. Large
dot marks the starting point of the particle.}
\label{f1}
\end{figure}

Fine adjustments of the initial velocity lead to a highly harmonious trajectory
(Fig.2), which looks like a loop orbit in a potential of a single bar
(see e.g. Fig.3.7a in \cite{bt87}). This is only a formal similarity, but
understanding it will let us find out what kind of orbit we see in Fig.2. The 
loop orbit in a single bar forms when a particle oscillates around a closed 
periodic orbit. Therefore two frequencies are involved: the frequency
of the free oscillation, and the forcing frequency of the bar. On the other
hand, the Fourier transform of the trajectory from Fig.2 shows two sharp
peaks at frequencies equal to the forcing frequencies of the two bars (Fig.3).
Thus the trajectory from Fig.2 also has two frequencies: this time these are 
the forcing frequencies from the two bars, while the free oscillation is 
absent. This is how the solution in the linear approximation (\S2.2) was
constructed. We conclude that in both the linear (epicyclic approximation) 
case and the general case we are dealing with doubly periodic orbits
in an oscillating potential of a double bar, with frequencies equal 
to the forcing frequencies of the bars. 
In the epicyclic approximation, these orbits have a nice feature that 
particles following them populate loops: closed curves that return to 
their original shape and position at every alignment of the bars. One may 
therefore expect that also in the general case these particles gather on loops.

\begin{figure}[t]
\vspace{-8mm}
\begin{center}
\includegraphics[width=12cm]{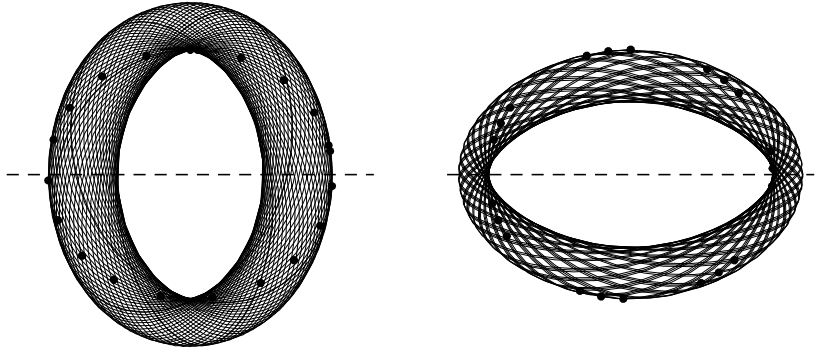}
\end{center}
\vspace{-6.5cm}
\caption[]{A doubly periodic orbit in the doubly barred potential,
followed for 20 relative periods of the bars, and written in the frame 
corotating with the big bar (left), and the small bar 
(right). The long axis of each bar is marked by the dashed line. Dots mark
positions of the particle at every alignment of the bars.}
\label{f2}
\end{figure}

\begin{figure}[b]
\vspace{-5mm}
\begin{center}
\includegraphics[width=8cm]{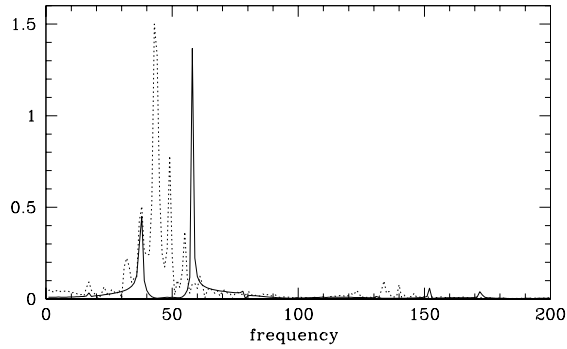}
\end{center}
\vspace{-3.5cm}
\caption[]{Fourier transforms of the trajectories from right panels of Fig.1 
(dotted line) and from Fig.2 (solid line). The peaks in the solid line are 
related to the forcing frequencies of the bars, and the peaks in the dotted 
line are not.}
\label{f3}
\end{figure}

If in the general case particles on doubly periodic orbits form a loop,
one can construct it by writing positions of a particle on such an orbit 
every time the bars align. These positions are the initial conditions for 
particles forming the loop, because after every alignment, the $n^{th}$ 
particle generated in this way takes the position of particle $n+1$. The 
first 20 positions of a particle on a doubly periodic orbit are overplotted 
in Fig.2. They indeed seem to be arranged on a closed ellipse-like curve; the 
shape of this curve varies in time, but it returns to where it started at
every alignment of the bars (Fig.4). This construction shows that in the 
general case particles on doubly periodic orbits also form loops. Note that
positions of particles on other orbits, which involve free oscillations,
when written at every alignment of the bars, densely populate some 
two-dimensional section of the plane, and do not gather on any curve.
It is extremely useful for the investigation of the orbital structure
in double bars that the appearance of the loop is frame-independent.
Loops provide an efficient way to classify doubly periodic orbits,
which has been hampered so far by the dependence of the last ones on 
the reference frame.

\begin{figure}[t]
\vspace{-2.8cm}
\begin{center}
\includegraphics[width=10cm]{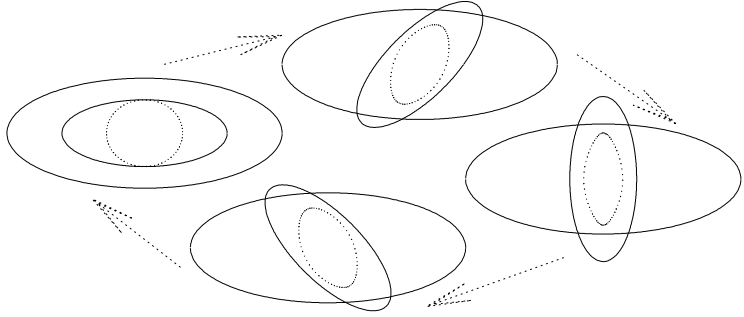}
\end{center}
\vspace{-2.8cm}
\caption[]{Evolution of the loop from Fig.2 during one relative period of the 
bars. The bars, outlined with solid lines, rotate counterclockwise. The loop 
is made 
out of points that represent separate particles on doubly periodic orbits.}
\label{f4}
\end{figure}

It turns out that doubly periodic orbits play crucial role in providing
orbital support for the pulsating potential of double bars. No closed
periodic orbits have been proposed as candidates for the backbone of
such a potential. If in a given potential of two bars there are loops that
follow the inner bar, and other loops that follow the outer bar, then
one may expect that such a potential is dynamically possible. An example
of such a potential has been constructed in \cite{m+s00}. The loop from Fig.4 
does not follow either bar in its motion, and therefore it is unlikely that 
it supports the assumed potential. It can be shown that in that potential, 
there are no loops which could support the two bars. Thus that potential 
is not self-consistent. This example shows how efficient is the loop approach
in rejecting hypothetical doubly barred systems that have no orbital support.

\begin{figure}[t]
%\vspace{-5mm}
\begin{center}
\includegraphics[width=14cm]{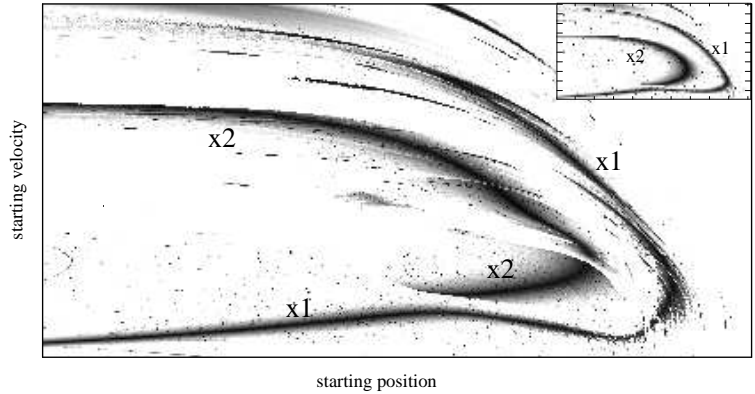}
\end{center}
\vspace{-12.5cm}
\caption[]{The width of the ring formed by particles trapped around loops 
in Model 2 from \cite{m+s00} as a function of the particle's position along 
the minor axis of the aligned bars, and of its velocity (perpendicular to 
this axis). Darker color means smaller width. In the insert, the same is 
shown for rings around closed periodic orbits in a single bar (same model, 
but inner bar axisymmetric). Regions related to the $x_1$ and $x_2$ orbits, 
and to the loops originating from them, are marked.}
\label{f5}
\end{figure}

Doubly periodic orbits in double bars are surrounded by regular orbits 
in the same way as are the closed periodic orbits in a single bar. In both
cases, the trapped regular orbits oscillate around the parent orbit. The
trajectory from the right panels of Fig.1 is an example of a regular orbit
that is trapped around the doubly periodic orbit from Fig.2. How much of the 
phase space in double bars is occupied by orbits trapped around doubly 
periodic orbits? It can be examined by launching a particle from e.g. the minor
axis of the bar, in the direction perpendicular to this axis, when the bars 
are aligned. If the particle is trapped, its positions at every alignment of 
the two bars lie within a ring containing the loop. The width of this ring
depends on the particle's position along the minor axis, and on its velocity.
It is displayed in Fig.5 for the potential of Model 2 defined in 
\cite{m+s00}. Two stripes of low width appear on the diagram, which correspond
to the $x_1$ and $x_2$ orbits in a single bar \cite{c+p80} (displayed in the
insert). Thus in double 
bars there are doubly periodic orbits that correspond to closed periodic 
orbits in single bars. There are possible regions of chaos in double bars 
(white stripes in Fig.5), but overall loops in double bars and periodic 
orbits in single bars trap similar volumes of phase-space around them.

\section{Conclusions}
In a potential of two independently rotating bars, a large fraction of phase 
space can be occupied by trajectories trapped around parent regular orbits.
These orbits are doubly periodic, with the two periods corresponding to the 
forcing frequencies of the two bars, but they do not close in any reference 
frame. Like particle trajectories oscillating around closed periodic orbits 
in a single bar, particle trajectories in double bars oscillate around the 
doubly periodic parent orbits. The structure of the parent regular orbits can 
be mapped using the loop approach, which allows us to single out dynamically 
possible double bars.

\vspace{1cm}

{\bf Acknowledgments.} The concept of the loop as the organized form of 
motion in double bars benefits from the insight of Linda Sparke. I thank Lia 
Athanassoula for our collaboration that lead to this paper, and Peter Erwin 
for comments on the manuscript.

\end{document}